\documentclass[aps,pre,showpacs,amsmath,amssymb,longbibliography]{revtex4-1}
\usepackage{amsmath,amssymb,latexsym,epsfig,graphicx,epsf,bm}
\usepackage{graphicx}
\usepackage{float}
\newcommand{\bF}{\bold F}

\newcommand{\bR}{\bold R}
\newcommand{\br}{\bold r}
\newcommand{\tbr}{\tilde{\bold r}}

\newcommand{\tbR}{{\tilde{\bold R}}}
\renewcommand{\tt}{\tilde{t}}

\newcommand{\be}{\begin{equation}}
\newcommand{\ee}{\end{equation}}
\newcommand{\fig}[1]{Fig.~\ref{#1}}
\newcommand{\Fig}[1]{Figure~\ref{#1}}
\newcommand{\sect}[1]{Sec.~\ref{#1}}
\newcommand{\Sec}[1]{Section~\ref{#1}}
\newcommand{\eq}[1]{Eq.~(\ref{#1})} 

\newcommand{\Sex}{{S}_{\rm ex}}
\newcommand{\Tc}{{T}_{\rm conf}}

\newcommand{\Teq}{{T}_{\rm eq}}
\newcommand{\Ts}{{T}_{\rm s}}

\newcommand{\bo}{\mathbf o}
\newcommand{\bxi}{\boldsymbol{\xi}}

\begin{document}
	\title{Configurational temperature in active matter. II. Quantifying the deviation from thermal equilibrium}
	\date{\today}
	\author{Shibu Saw}\email{shibus@ruc.dk}
	\affiliation{\textit{Glass and Time}, IMFUFA, Department of Science and Environment, Roskilde University, P.O. Box 260, DK-4000 Roskilde, Denmark}
	\author{Lorenzo Costigliola}\email{lorenzo.costigliola@gmail.com}
	\affiliation{\textit{Glass and Time}, IMFUFA, Department of Science and Environment, Roskilde University, P.O. Box 260, DK-4000 Roskilde, Denmark}	\email{lorenzo.costigliola@gmail.com}
	\author{Jeppe C. Dyre}\email{dyre@ruc.dk}
	\affiliation{\textit{Glass and Time}, IMFUFA, Department of Science and Environment, Roskilde University, P.O. Box 260, DK-4000 Roskilde, Denmark}

\begin{abstract}
This paper suggests using the configurational temperature $\Tc$ for quantifying how far an active-matter system is from thermal equilibrium. We measure this ``distance'' by the ratio of the systemic temperature $\Ts$ to $\Tc$, where $\Ts$ is the canonical-ensemble temperature for which the average potential energy is equal to that of the active-matter system. $\Tc$ is ``local'' in the sense that it is the average of a function, which only depends on how the potential energy varies in the vicinity of a given configuration; in contrast $\Ts$ is a global quantity. The quantity $\Ts/\Tc$ is straightforward to evaluate in a computer simulation; equilibrium simulations in conjunction with a single steady-state active-matter configuration are enough to determine $\Ts/\Tc$. We validate the suggestion that $\Ts/\Tc$ quantifies the deviation from thermal equilibrium by data for the radial distribution function of 3d Kob-Andersen and 2d Yukawa active-matter models with active Ornstein-Uhlenbeck and active Brownian Particle dynamics. Moreover, we show that $\Ts/\Tc$, structure, and dynamics of the homogeneous phase are all approximately invariant along the motility-induced phase separation (MIPS) boundary in the phase diagram of the 2d Yukawa model. The measure $\Ts/\Tc$ is not limited to active matter; it can be used for quantifying how far any system involving a potential-energy function, e.g., a driven Hamiltonian system, is from thermal equilibrium. 
\end{abstract}

\maketitle

\section{Introduction}\label{I}

Temperature is fundamental in thermodynamics and statistical mechanics. Generalizations of the temperature concept to deal with out-of-equilibrium systems have been discussed in several publications, useful reviews of which are given in Refs. \onlinecite{cas03,pow05,leu09,pug17,zha19}. Non-equilibrium temperatures generally attempt to relate the non-equilibrium system to the its thermal equilibrium properties. This paper and its companion \cite{saw23a}, henceforth referred to as Paper I, propose two applications of the so-called \textit{configurational temperature}  $\Tc$ \cite{LLstat,rug97,pow05,him19} to active-matter models, both of which are based on a different philosophy. Paper I showed that $\Tc$ defines an energy scale, which can be used for tracing out lines of approximately invariant physics of the 3d Kob-Andersen binary Lennard-Jones model with active Ornstein-Uhlenbeck dynamics. The present paper shows that a similar procedure applies for the 2d Yukawa model with active Brownian dynamics (ABP), after which we proceed to the main focus: using $\Tc$ for measuring how far an active-matter system is from thermal equilibrium.

For an ordinary Hamiltonian system in thermal equilibrium, the temperature $T$ is identical to the configurational temperature $\Tc$ that is defined \cite{rug97,pow05} as follows. If the system consists of $N$ particles with collective coordinate vector $\bR\equiv (\br_1,...,\br_N)$ and potential-energy function $U(\bR)$, one defines $k_B\Tc\equiv\langle(\nabla U)^2\rangle/\langle\nabla^2 U\rangle$. Here $k_B$ is the Boltzmann constant, $\nabla$ is the gradient operator, and the sharp brackets denote canonical-ensemble averages. It is straightforward to prove that $T=\Tc$ in equilibrium \cite{LLstat}, see, e.g., Paper I. Approaching the thermodynamic limit, the relative fluctuations of both the numerator and the denominator of $\Tc$ goes to zero. Thus if one defines an $\bR$-dependent configurational temperature by

\be\label{Tc_def}
k_B\Tc(\bR)
\,\equiv\,\frac{(\nabla U(\bR))^2}{\nabla^2 U(\bR)}\,,
\ee
the identity $\Tc(\bR)\cong T$ applies in thermal equilibrium in the sense that deviations vanish as $N\to\infty$. Because configurations with $\nabla^2 U(\bR)\leq 0$ become less likely as $N\to\infty$, the fact that \eq{Tc_def} is not defined for such configurations does not present a serious problem.

The derivation and justification of the configurational temperature $\Tc$ is based on the fact that the probability of configuration $\bR$ in the canonical ensemble is proportional to $\exp(-U(\bR)/k_BT)$ \cite{LLstat,pow05,saw23a}. This is irrelevant, however, for the property demonstrated in Paper I that $\Tc(\bR)$ may be used for tracing out lines of invariant structure and dynamics in the phase diagram of active-matter models that involve a potential-energy function obeying \textit{hidden scale invariance} \cite{dyr14}. This is the symmetry that the ordering of configurations according to their potential energy at a given density is maintained if these are scaled uniformly to a different density. Hidden scale invariance applies to a good approximation for many well-known potentials, e.g., systems defined by the Lennard-Jones and Yukawa interactions, density-functional derived atomic interactions, and simple molecular models \cite{IV,ing12b,sch14,hum15,dyr18a}.

This paper proposes an application of $\Tc$ to active-matter models, which addresses the problem of quantifying how far a system is from ordinary canonical-ensemble thermal equilibrium. This question is important because only if the system in question is close to thermal equilibrium, does it make good sense to refer to the temperature of the corresponding canonical-ensemble equilibrium system as a characteristic of the active-matter system. As discussed in the next section, the ratio between the global ``systemic'' temperature $\Ts$ and the ``local'' temperature $\Tc$ provides such a measure. \Sec{III} sets the stage by detailing one example, the 2d Yukawa model with active Brownian particle dynamics. \Sec{quant} presents data for the radial distribution function of Kob-Andersen and 2d Yukawa active-matter models, confirming that when $\Ts/\Tc$ is close to unity, the structure is close to that of thermal equilibrium. Also, \sect{quant} evaluates a standard entropy-production-based measure of deviations from thermal equilibrium and compared to the proposed new measure. \Sec{MIPS} shows that the new measure is roughly constant along the motility-induced phase-separation line, which is consistent with the reasonable assumption that all state points close to this line in the non-MIPS phase are equally far from equilibrium. Finally, \sect{out} summarizes Papers I and II.

\section{How far is a given active-matter system from thermal equilibrium?}\label{II}

The investigations of Papers I and II are limited to active-matter point-particle models characterized by a potential-energy function. Quantifying the degree of non-equilibrium is usually done by calculating some form of dissipation (entropy production). The idea is that since the entropy production is zero in thermal equilibrium, this quantity measures how far a given system is from thermal equilibrium \cite{fod16,fle20,byr22}. Such measures can be applied to both active-matter models and driven Hamiltonian systems. A fundamental issue with these measures is the following. Using a quantity that goes to zero in some limit to quantify the degree of deviation from that limit does not in an obvious way make possible the identification of when deviations from equilibrium are to be regarded as ``large''. If deviations from thermal equilibrium are instead quantified by means of a quantity that goes to \textit{unity} in the equilibrium limit, deviations from equilibrium are ``small'' whenever that quantity does not deviate substantially from unity and ``large'' otherwise. 

The configurational temperature is local in the sense that when regarded as a function of $\bR$, it only depends on how the potential energy $U(\bR)$ varies in the immediate surroundings. Note that ``local'' here refers to the 2N or 3N dimensional configuration space, not to the two- or three-dimensional space in which the active particles move. This locality means that by evaluating $\Tc$ for a passive system's configuration at a given time, one cannot determine whether the system is in thermal equilibrium corresponding to the temperature $T=\Tc(\bR)$. For instance, for an aging glass annealed at temperature $T$, already after a time on the phonon scale does $\Tc(\bR)\cong T$ apply, i.e., long before equilibrium has been reached \cite{pow05}. A completely different, global temperature concept is the systemic temperature $\Ts$. This quantity was introduced for generalizing isomorph theory to out-of-equilibrium conditions \cite{dyr20}, but $\Ts$ may be introduced for any system as the equilibrium canonical-ensemble temperature of the Hamiltonian system at the same density and average potential energy as that of the out-of-equilibrium system. In thermal equilibrium one has $\Tc=\Ts=T$. 

The idea is to use the ratio of global to local temperature, $\Ts/\Tc$, for quantifying how far an active-matter system is from thermal equilibrium. We showed in Paper I that the ratio $\Ts/\Tc$ is predicted to be constant along active-matter isomorphs. Since structure and dynamics are also invariant along both active-matter isomorphs and the corresponding Hamiltonian-system isomorphs, it is consistent to assume that $\Ts/\Tc$ measures how far the system is from thermal equilibrium.

\section{The Yukawa Active Brownian-particle model in two dimensions}\label{III}

This section details the ABP model in two dimensions based on the single-component Yukawa pair potential \cite{yuk35,mea21},

\be\label{yuk}
v(r)
\,=\,\frac{Q^2\,\sigma}{r}\,e^{-r/(\lambda\sigma)}\,.
\ee
This potential obey hidden scale invariance \cite{dyr14,vel15,tol19}, so a procedure for identifying active-matter isomorphs analogous to that introduced in Paper I for the active Ornstein-Uhlenbeck particle (AOUP) model should apply here, as well. The idea is that $\Ts/\Tc$, as mentioned, is predicted to be invariant along active-matter isomorphs where the deviations from thermal equilibrium are also expected to be invariant. 

If $\br_i$ is the position vector of particle $i$, the ABP equations of motion in two dimensions are

\be\label{ABP_EOM2}
\dot{\br}_i
\,=\,\mu\bF_i+\bxi_i(t)+v_0\, \bo_i(t)\,.
\ee
Here, $\mu$ is the mobility, $\bF_i(\bR)=-\nabla_i U(\bR)$ is the force on particle $i$, $\bxi_i(t)$ is a Gaussian random white-noise vector, $v_0$ is a constant velocity, and $\bo_i(t)=(\cos(\theta_i(t)),\sin(\theta_i(t)))$ is a stochastic unit vector. The direction vector angle $\theta_i(t)$ is controlled by a white Gaussian noise of magnitude $D_r$,

\be\label{oi_noise}
\langle\dot\theta_i(t)\dot\theta_j(t')\rangle
\,=\,2D_r\delta_{ij}\,\delta(t-t')\,,
\ee
and the white-noise vector has magnitude $D_t$,

\be\label{xi_noise2}
\langle\bxi_i^\alpha(t)\bxi_j^\beta(t')\rangle
\,=\,2D_t\delta_{ij}\delta_{\alpha\beta}\delta(t-t')\,.
\ee

The ABP model has four parameters. Regarding $\mu$ as a system-specific constant, the dimensionless versions of the three other parameters must be constant in order to have invariant physics when the density is changed. Following the procedure of Sec. III of Paper I, we take as length unit $l_0=\rho^{-1/2}$ (the exponent is $-1/2$ and not $-1/3$ as in Paper I because the model here is two-dimensional) and as time unit $t_0=1/D_r$, and write the equation of motion in terms of the corresponding reduced variables. Substituting $\br_i=\rho^{-1/2}\tbr_i$ and $t=\tt/D_r$ into \eq{ABP_EOM2} and making use of Eq. (8) of Paper I and the definition of the systemic temperature $\Ts$ \cite{dyr20} in which $\Sex(\bR)$ is the microscopic excess-entropy function \cite{sch14,dyr20},

\be\label{Tseq}
\Ts(\bR)
\,\equiv\,\Teq(\rho,\Sex(\bR))=\Teq(\rho,U(\bR))\,,
\ee
we get

\be\label{ABP_EOM3}
\dot{\tbr}_i
\,=\,-\mu \rho(\Ts/D_r)\tilde{\nabla}_i\Sex(\tbR)+\tilde{\bxi}_i(t)+\tilde{v}_0\, \bo_i(t)\,.
\ee
Here $\tilde{v}_0=(\rho^{1/2}/D_r)v_0$, $\tilde\bxi_i=(\rho^{1/2}/D_r)\bxi_i$, $\Ts$ is brief for $\Ts(\bR)$,

\be\label{xi_noise3}
\langle\tilde{\bxi}_i^\alpha(t)\tilde{\bxi}_j^\beta(t')\rangle
\,=\,2\rho(D_t/D_r)\delta_{ij}\delta_{\alpha\beta}\delta(\tt-\tt')\,,
\ee
and dots now mark the derivative with respect to $\tt$,

\be\label{oi_noise2}
\langle\dot\theta_i(t)\dot\theta_j(t')\rangle
\,=\,2\delta_{ij}\,\delta(\tt-\tt')\,.
\ee
These equations are invariant under a change of density if $\mu\rho\Ts/D_r$, $\rho D_t/D_r$, and $\tilde{v}_0$ are kept constant. Since $\mu$ is a (system-specific) constant, this implies (where the subscript zero refers to a reference state of density $\rho_0$ and $\Ts(\rho)\equiv\Teq(\rho,\Sex(\tbR))$ can be used instead of $\Ts(\bR)$ because fluctuations go to zero in the thermodynamic limit)

\begin{eqnarray}\label{ABP_param}
D_r&\,=\,&D_{r,0}\,\frac{\rho}{\rho_0}\,\frac{\Ts(\rho)}{\Ts(\rho_0)}\nonumber\\
D_t&\,=\,&D_{t,0}\,\frac{\Ts(\rho)}{\Ts(\rho_0)}\\
v_{0}&\,=\,&v_{0,0}\,\left(\frac{\rho}{\rho_0}\right)^{1/2}\frac{\Ts(\rho)}{\Ts(\rho_0)}
\nonumber\,.
\end{eqnarray}
By the same argument as in Sec. III of Paper I one can here replace the $\Ts$ ratios by $\Tc$ ratios, leading to

\begin{eqnarray}\label{ABP_param2}
D_r&\,=\,&D_{r,0}\,\frac{\rho}{\rho_0}\,\frac{\Tc\left((\rho_0/\rho)^{1/2}\bR_0\right)}{\Tc(\bR_0)}\nonumber\\
D_t&\,=\,&D_{t,0}\,\frac{\Tc\left((\rho_0/\rho)^{1/2}\bR_0\right)}{\Tc(\bR_0)}\\
v_{0}&\,=\,&v_{0,0}\,\left(\frac{\rho}{\rho_0}\right)^{1/2}\frac{\Tc\left((\rho_0/\rho)^{1/2}\bR_0\right)}{\Tc(\bR_0)}\nonumber\,.
\end{eqnarray}
In passing we note that while the Peclet number $v_0/\sqrt{2D_rD_t}$ \cite{bec16,hec21} is invariant along the active-matter isomorph, this requirement is not enough to determine how to scale the model parameters -- thus Peclet-number invariance is a necessary, but not sufficient condition for identifying an active-matter isomorph.

\begin{table}[H]
	\begin{center}
		\begin{tabular}{|l|c|c|c|r|}\hline
			$\rho$    &     $D_r$       &   $D_t$       &   $v_0$     &     $\Tc$        \\ \hline
			$1.0$     &     $3.000$     &   $1.000$     &   $25.00$   &     $1.489$      \\ \hline
			$1.5$     &     $12.37$     &   $2.750$     &   $84.20$   &     $4.093$      \\ \hline
			$2.0$     &     $30.43$     &   $5.072$     &   $179.3$   &     $7.550$      \\ \hline
			$2.5$     &     $58.13$     &   $7.751$     &   $306.4$   &     $11.54$      \\ \hline
			$3.0$     &     $95.82$     &   $10.65$     &   $461.0$   &     $15.85$      \\ \hline
		\end{tabular}
		\caption{Values of $\rho$, $D_r$, $D_t$, $v_0$, and $\Tc$ along the active-matter isomorph of the 2d Yukawa ABP model determined by \eq{ABP_param2}. By means of \eq{Tc_def} the configurational temperature $\Tc(\rho)$ is determined from a single configuration $\bR_0$ scaled to density $\rho$.}
		\label{tab2}
	\end{center}
\end{table}

To validate the existence of active-matter isomorphs according to the above prediction we simulated $N=10000$ particles of the 2d Yukawa system with $Q=50$, $\lambda=0.16$, $\sigma = 1$ defining the length unit, and a cutoff at $4.5\sigma$. The time step used is given by $\Delta t = \Delta \tilde t (D_t/{v_0}^2)$, where $\Delta \tilde t = 0.0625$ so that $\Delta t = 0.0001$ at the reference state point defined by $(\rho, D_r, D_t, v_0) = (1.0, 3.0, 1.0, 25.0)$. The simulations were carried out on GPU cards using a home-made code. An active-matter isomorph was traced out for densities varying a factor of three using \eq{ABP_param2} for a configuration $\bR_0$ selected from a steady-state simulation at the reference state point. Table \ref{tab2} gives the parameters obtained from \eq{ABP_param2}.

\begin{figure}[h]
	\includegraphics[width=8cm]{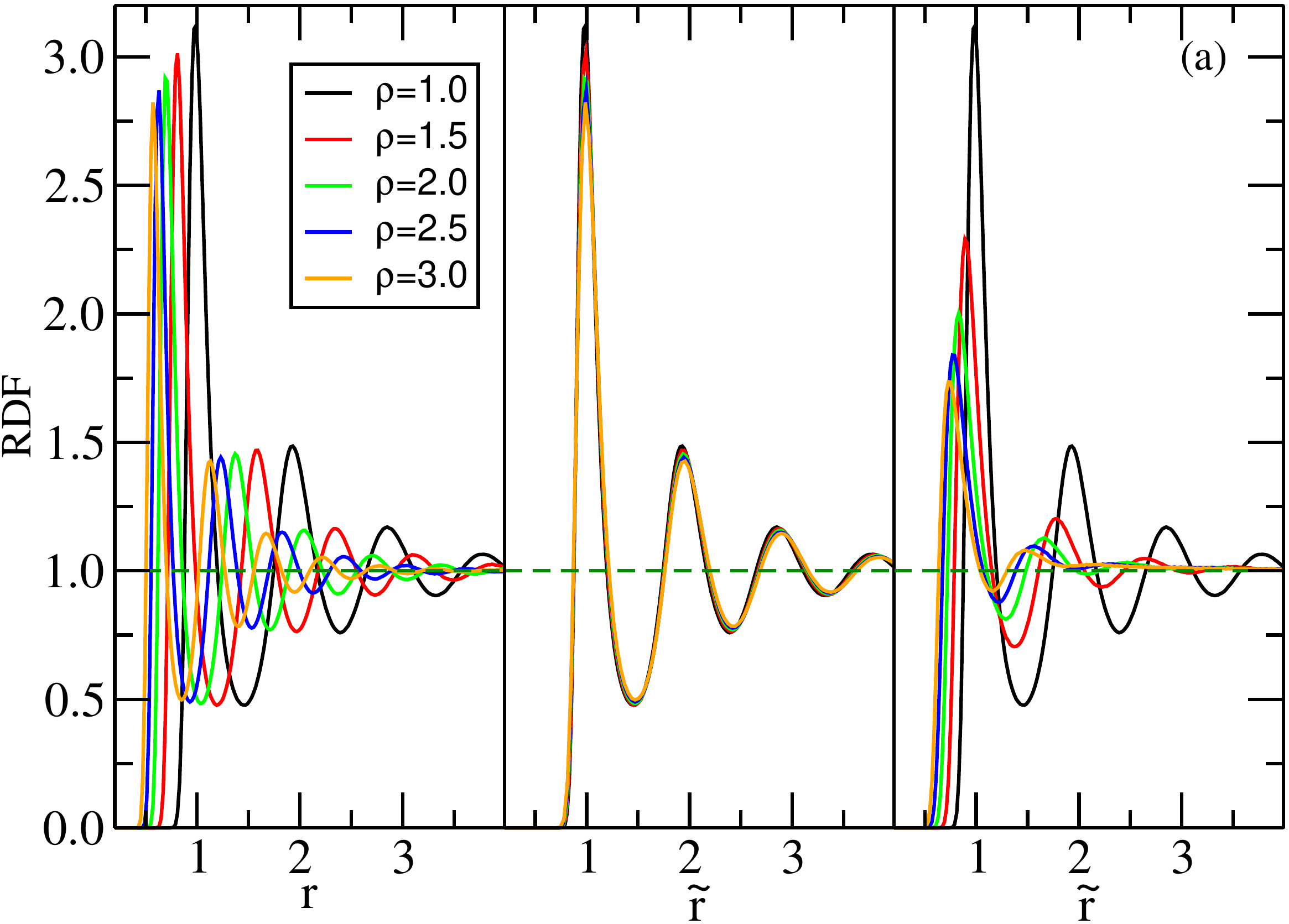}
	\includegraphics[width=8cm]{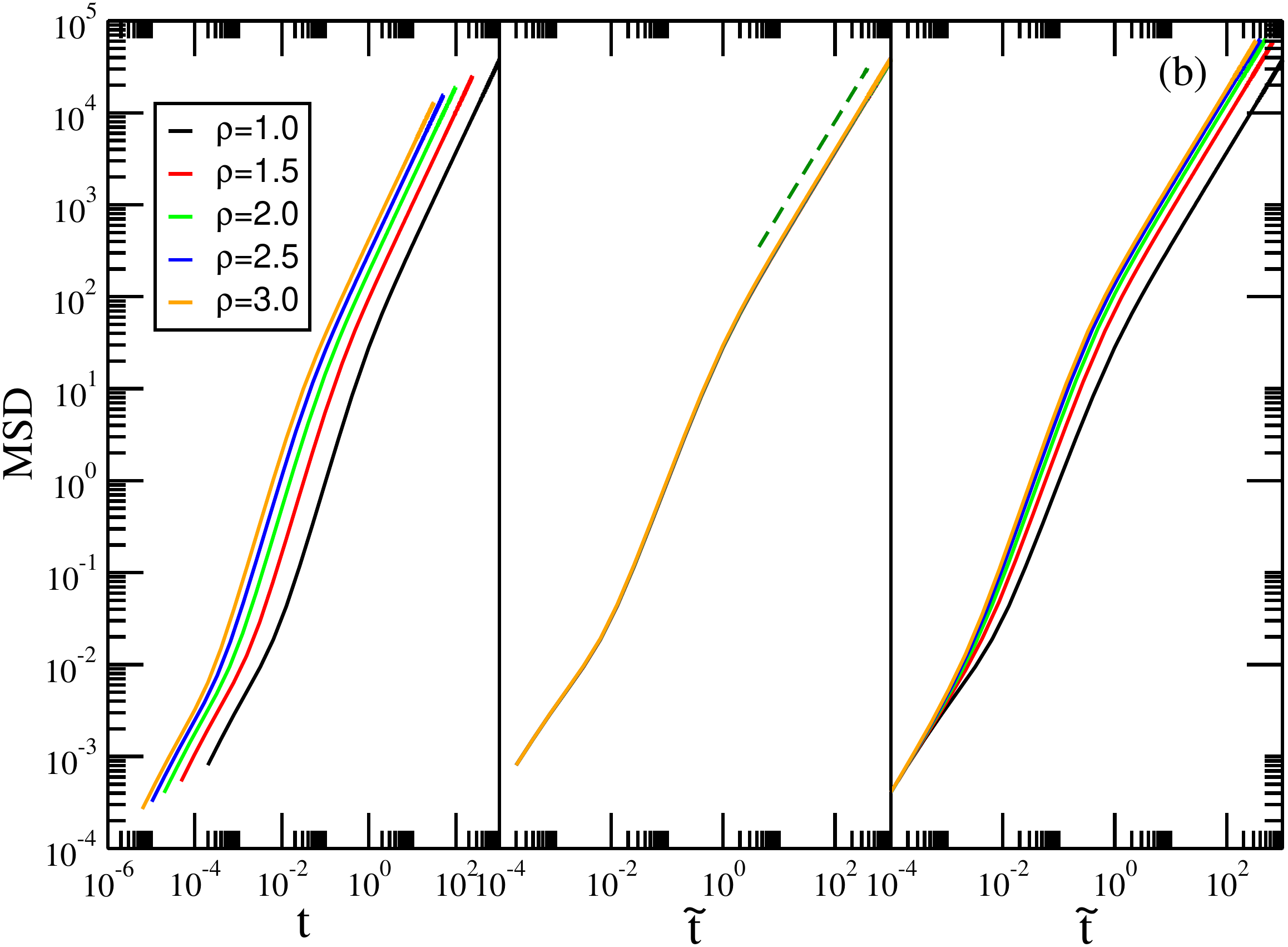}	
	\caption{\label{fig1} Structure and dynamics of the Yukawa ABP model in two dimensions.
	(a) The left panel shows the RDF as a function of the pair distance $r$ along the active-matter isomorph, the middle panel shows the same data in reduced units, and the right panel shows the reduced RDF for the same parameters (Table \ref{tab2}) at the reference density $\rho=1.0$. 
	(b) The left panel shows the MSD as a function of time $t$ along the active-matter isomorph, the middle panel shows the same data in reduced units where the dashed line marks slope unity, i.e., ordinary diffusion; the right panel shows the reduced MSD for the same parameters at the reference state-point density $\rho=1.0$.}
\end{figure}

\Fig{fig1}(a) shows the radial distribution function (RDF). The left two panels show the RDF along the active-matter isomorph as a function of $r$ and $\tilde{r}$, respectively. For comparison, the right panel shows the results for the same parameters at the reference state-point density $\rho=1.0$. We find a good invariance of the reduced RDF along the active-matter isomorph. The same applies for the reduced mean-square displacement (MSD) shown in (b).

\section{Deviations from thermal equilibrium quantified by $\Ts/\Tc$}\label{quant}

\begin{figure}[htbp!]
	\includegraphics[width=6cm]{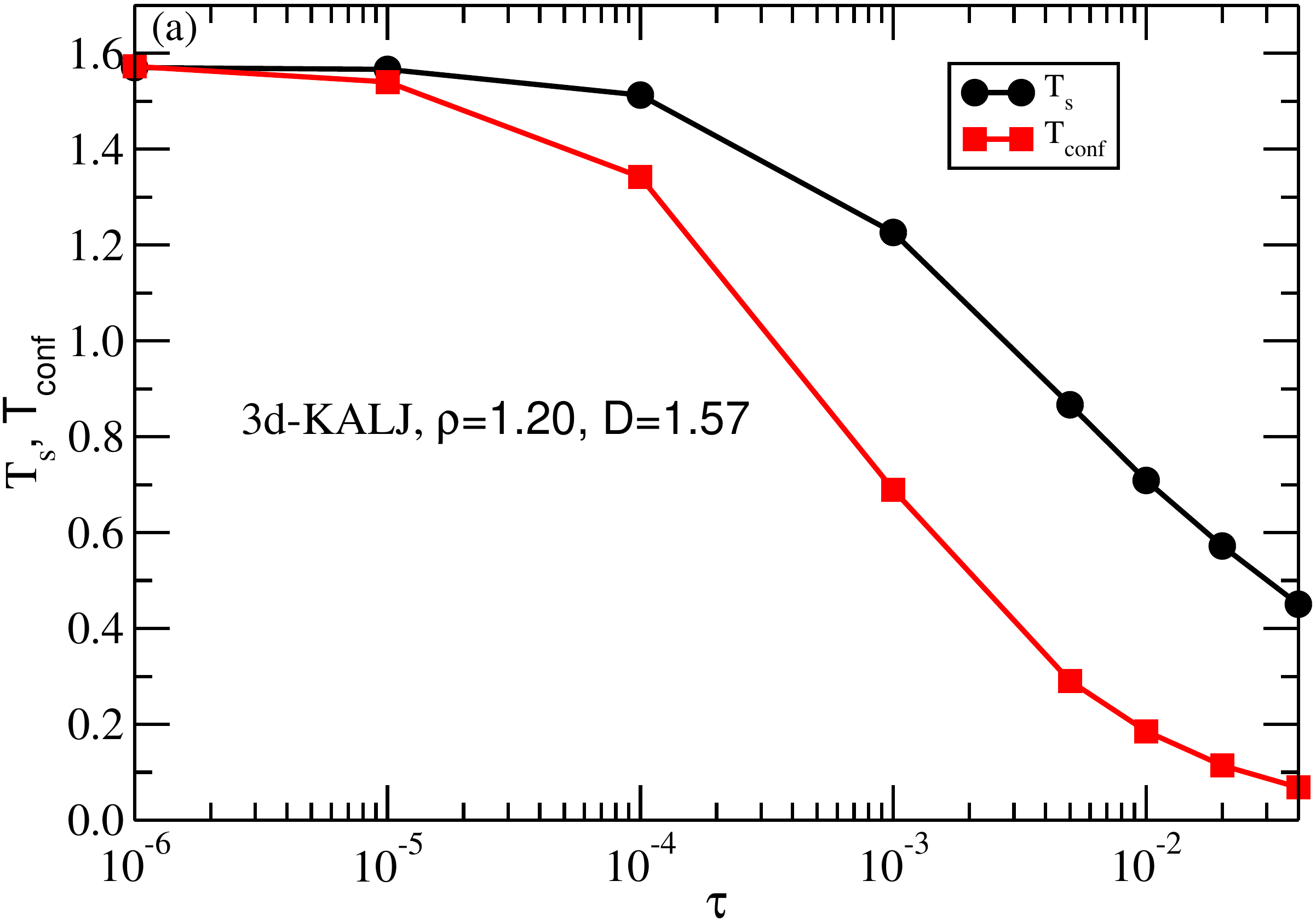}
	\includegraphics[width=6cm]{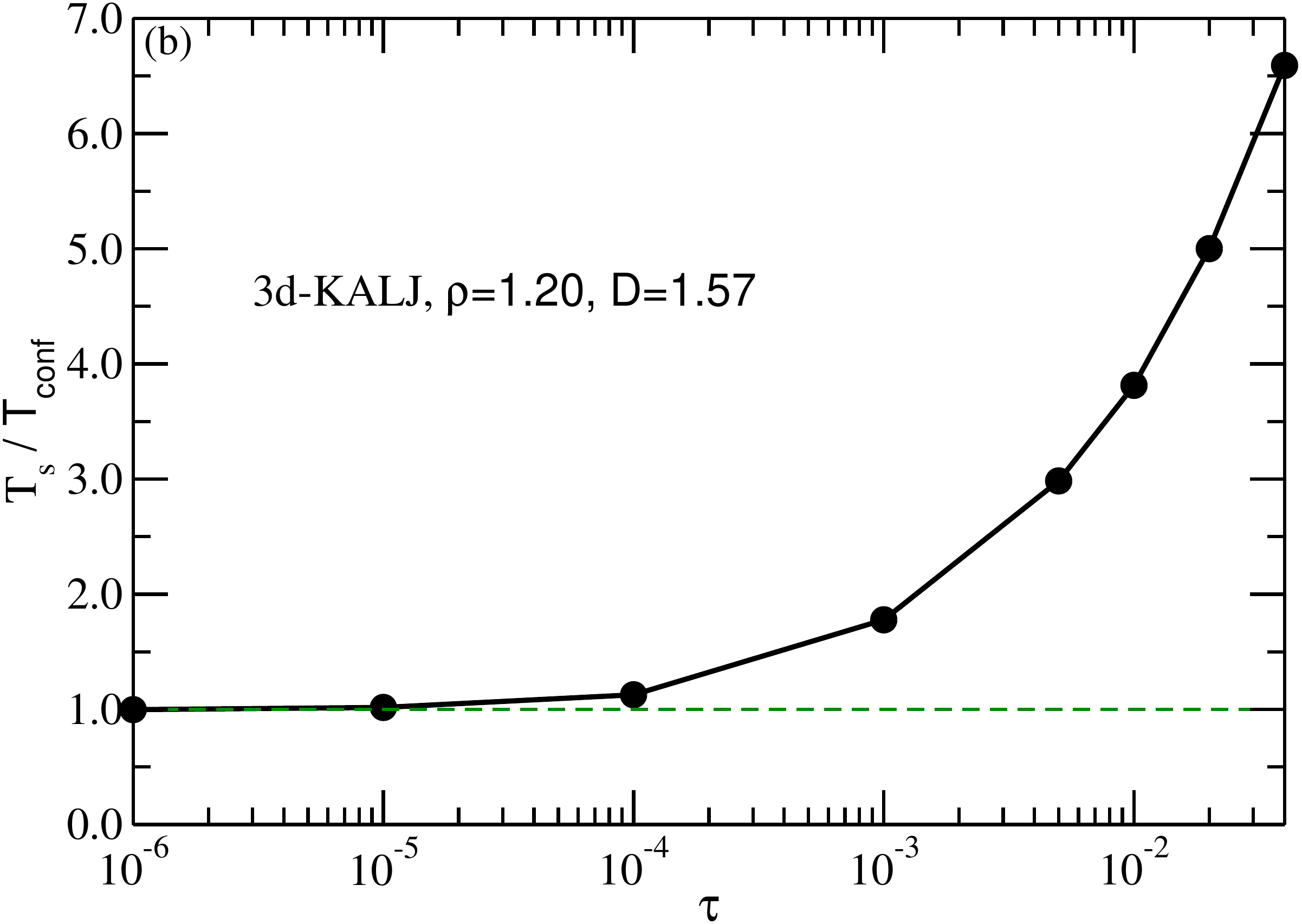}
	\includegraphics[width=6cm]{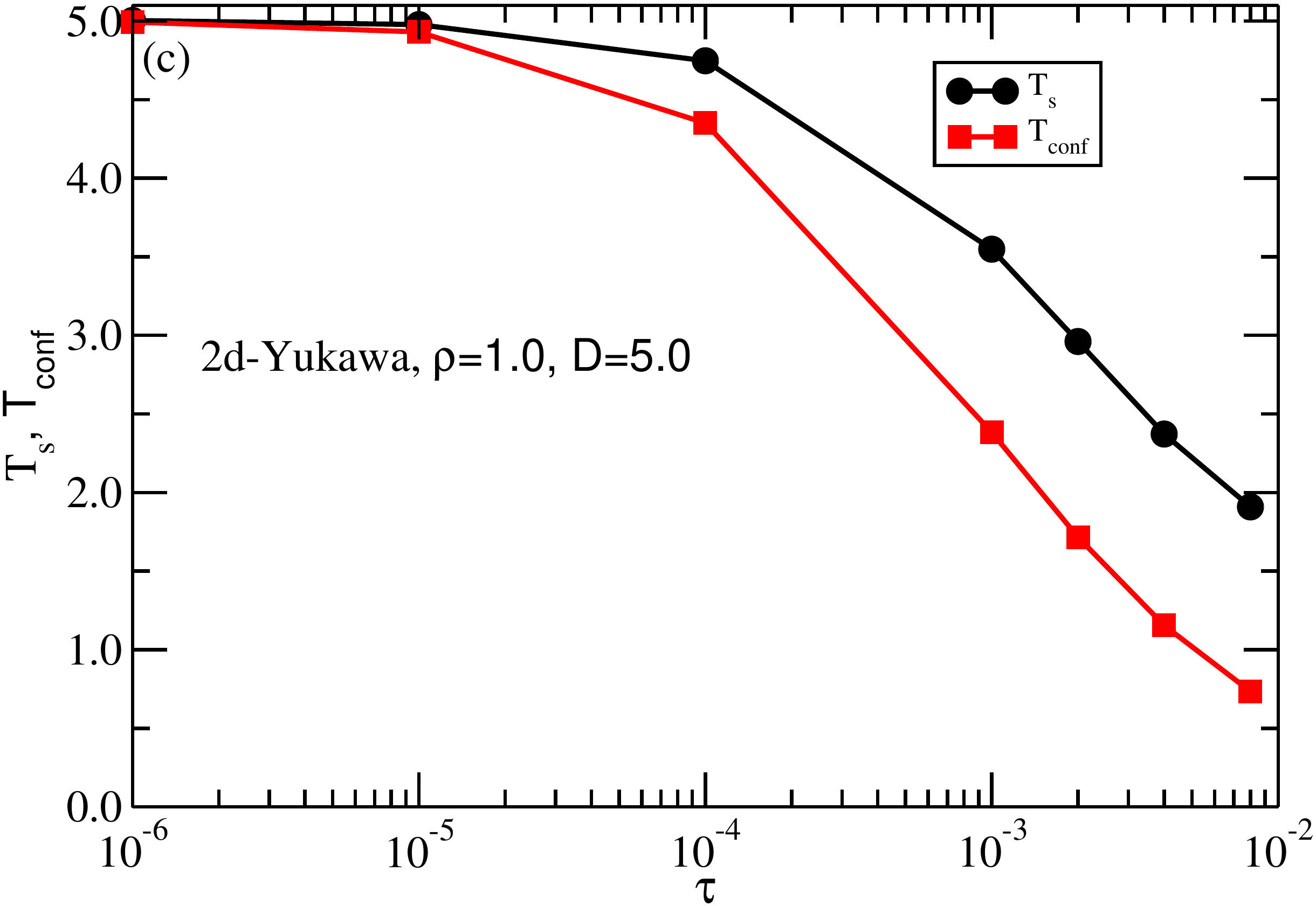}
	\includegraphics[width=6cm]{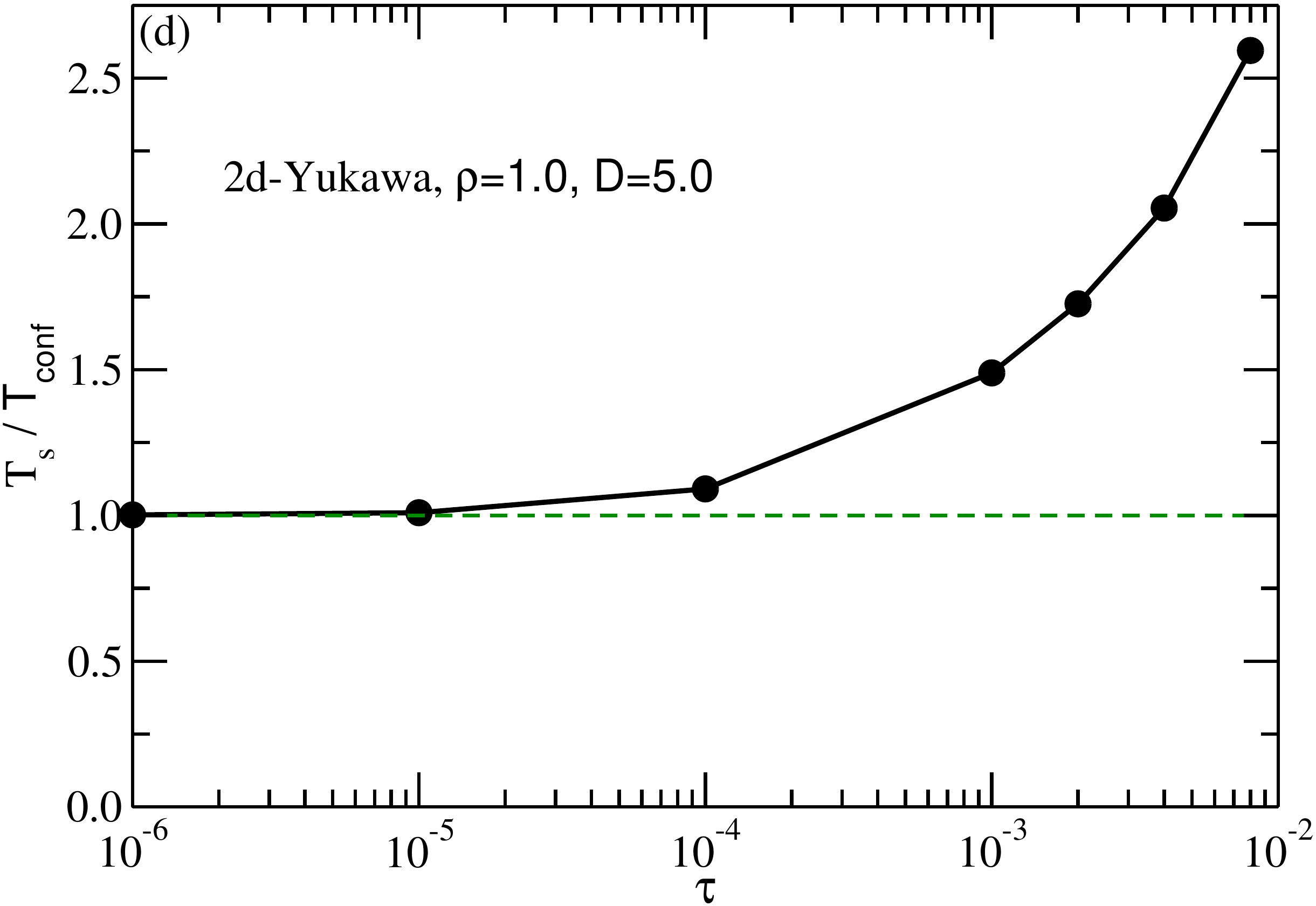}
	\includegraphics[width=6cm]{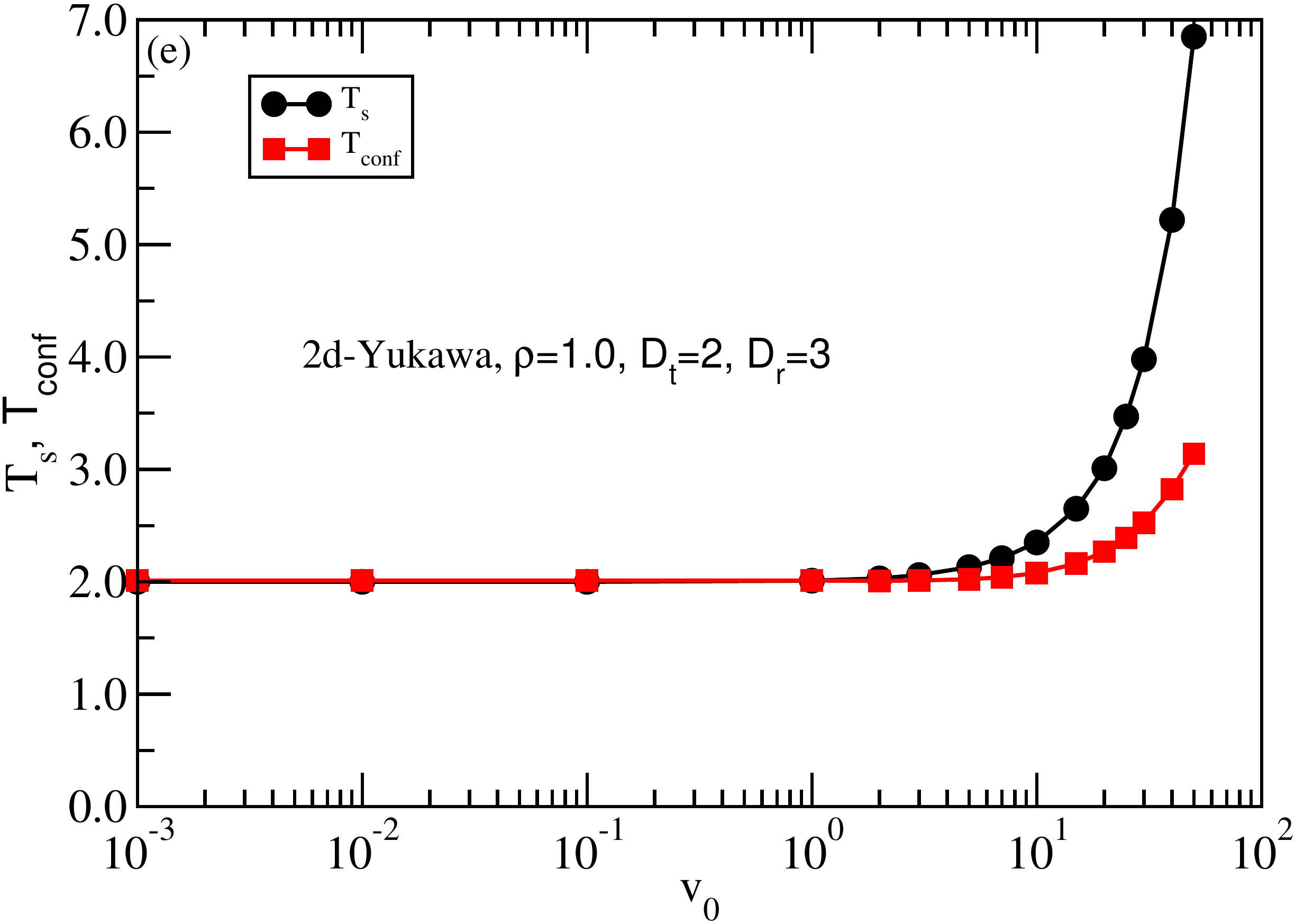}
	\includegraphics[width=6cm]{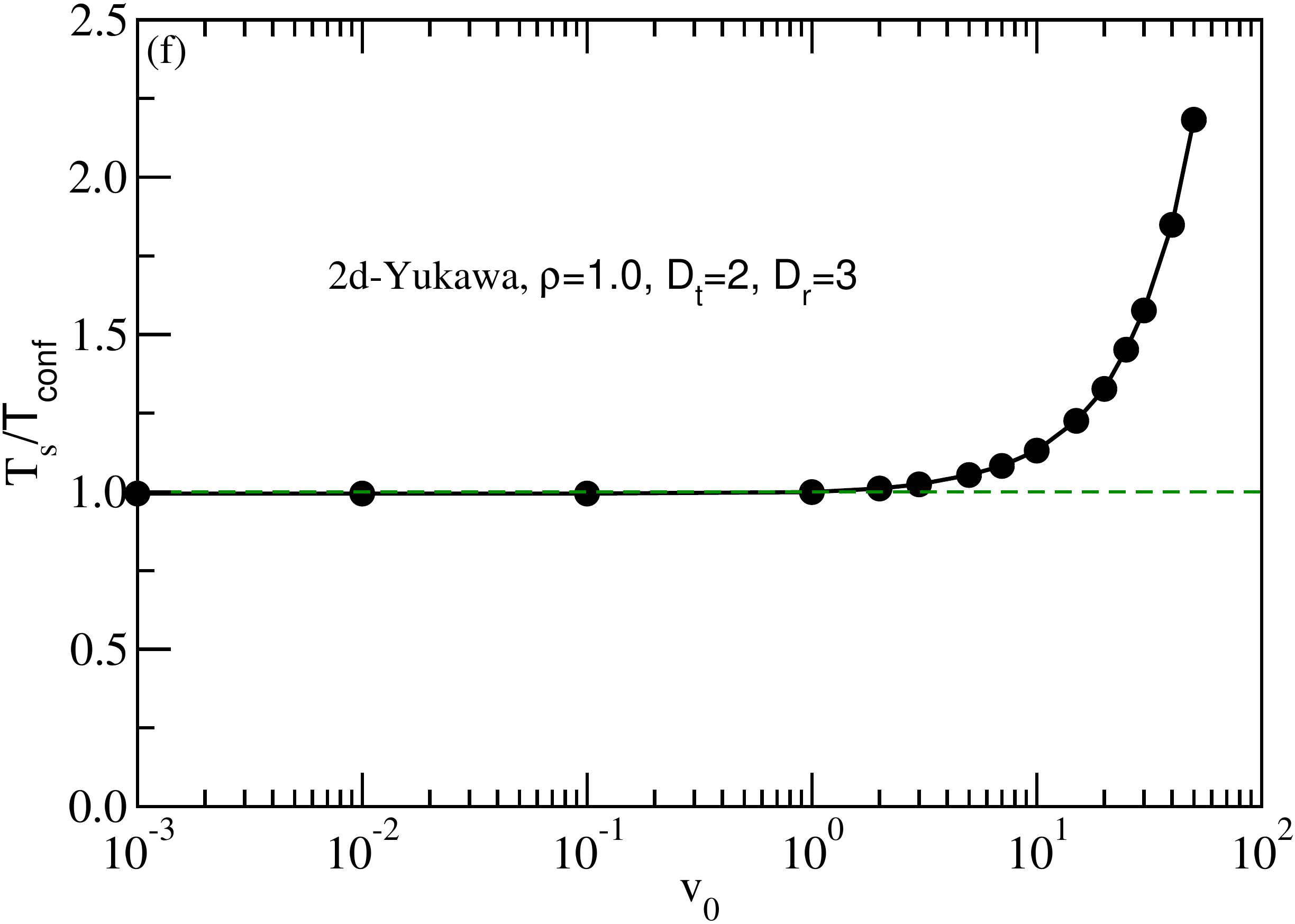}
	\caption{\label{fig2} Determination of the ratio of systemic to configurational temperature, $\Ts/\Tc$, quantifying how far an active-matter system is from thermal equilibrium. 
	(a) shows data for $\Ts$ and $\Tc$ for the 3d Kob-Andersen AOUP model (Paper I, \cite{saw23a}) as functions of $\tau$ with the remaining model parameters kept fixed.
	(b) shows $\Ts/\Tc$ for the same data. For $\tau$ values around $10^{-4}$ the system begins to move away from equilibrium and for $\tau>10^{-3}$ significant deviations from equilibrium are predicted.
	(c) shows data for $\Ts$ and $\Tc$ for the 2d Yukawa AOUP model as functions of $\tau$ with the remaining model parameters kept fixed.
	(d) shows $\Ts/\Tc$ for the same data. For $\tau$ values above $10^{-4}$ the system starts to deviate from equilibrium.
	(e) shows data for $\Ts$ and $\Tc$ for the 2d Yukawa ABP model as functions of $v_0$ with the remaining model parameters kept fixed.
	(f) shows $\Ts/\Tc$ for the same data. For $v_0$ values around $10$ the system begins to move away from equilibrium.}
\end{figure}

\Fig{fig2} gives data for the systemic and configurational temperatures of different active-matter models, starting with the Kob-Andersen model studied in Paper I. \Fig{fig2}(a) shows the systemic temperature $\Ts$ (black symbols) and the configurational temperature $\Tc$ (red symbols) for the Kob-Andersen AOUP active-matter model as functions of the colored-noise correlation time $\tau$ for fixed values of the other model parameters. As mentioned, $\Ts$ is determined by identifying the equilibrium temperature at which the system for a standard MD simulation has the same average potential energy as the AOUP system. The system approaches an equilibrium system for $\tau\to 0$, corresponding to the canonical-ensemble temperature $T=1.6$. \Fig{fig2}(b) plots the ratio $\Ts/\Tc$. We see that for values of $\tau$ above $10^{-4}$, the system starts to move away from thermal equilibrium. \Fig{fig2}(c) shows $\Ts$ and $\Tc$ as functions of $\tau$ for the 2d Yukawa AOUP model for fixed values of the other model parameters. Both $\Ts$ and $\Tc$ converge to $5$ as $\tau \to 0$, confirming the fact that $T=5$ is the equilibrium Brownian-dynamics temperature corresponding to the parameters $D_t=5$, $\mu=1$. \Fig{fig2}(d) shows $\Ts/\Tc$ and we see that for $\tau$ above $10^{-4}$, the system begins to deviate from thermal equilibrium. \Fig{fig2}(e) and (f) show $\Ts$ and $\Tc$ and their ratio for the 2d Yukawa ABP model as functions of $v_0$ for fixed values of the other model parameters; here $v_0>10$ is the approximate criterion for deviations from equilibrium.

\newpage
\begin{figure}[htbp!]
	\includegraphics[width=12cm]{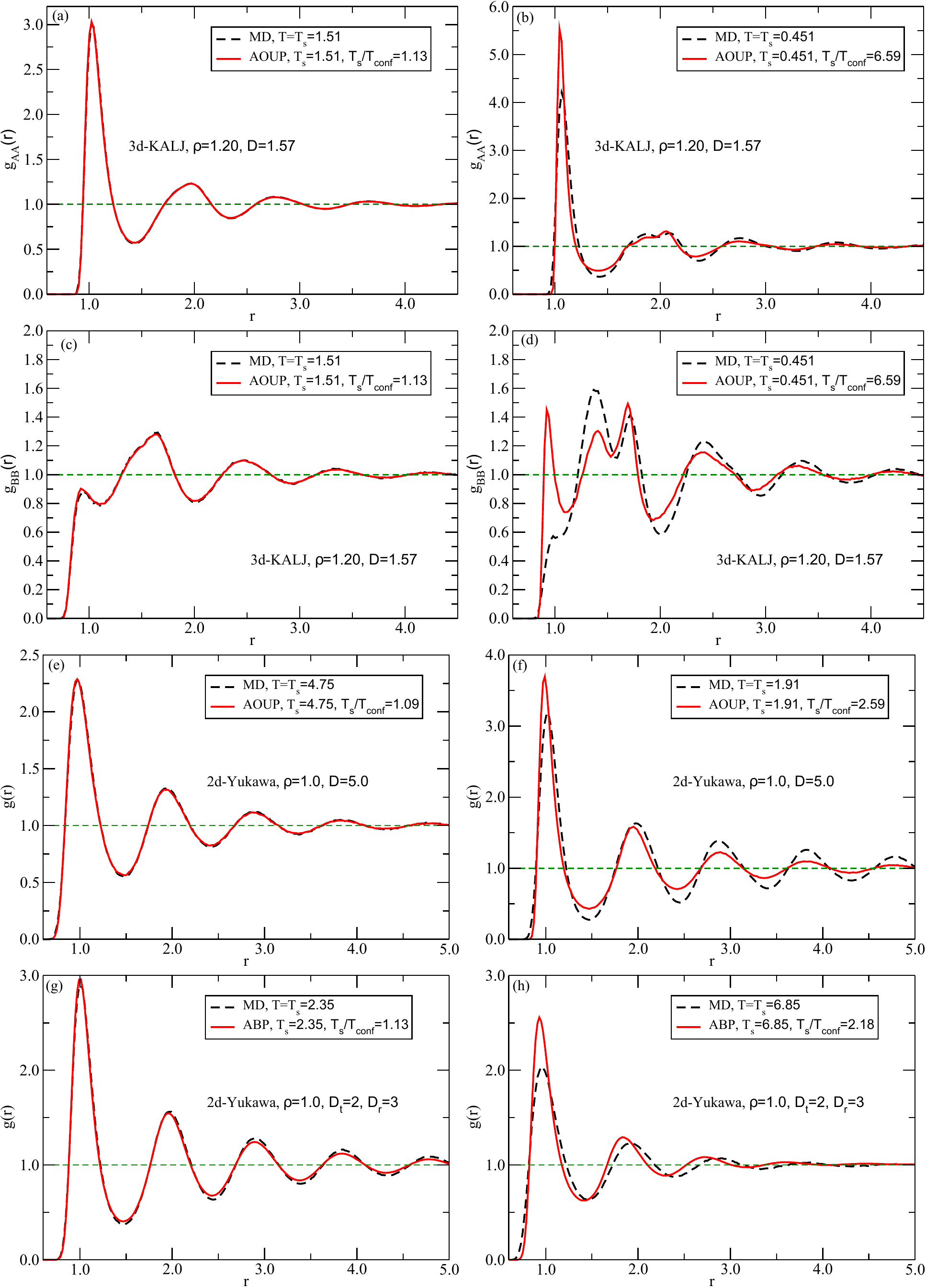}
	\caption{\label{fig2ex} RDFs of active-matter states predicted to be close to (left column) and not close to (right column) thermal equilibrium. The red curves are the active-matter data and the black dashed lines are the RDFs of the corresponding equilibrium system for $T=\Ts$.
	(a)-(d) show results for the AA and BB RDFs of the Kob-Andersen AOUP model for $\tau=10^{-4}$ and $\tau=4\cdot 10^{-2}$ (red curves) corresponding to $\Ts/\Tc=1.13$ and $\Ts/\Tc=6.59$.	
	(e) and (f) show results for the 2d Yukawa AOUP model at states with $\tau=10^{-4}$ and $\tau=8\cdot 10^{-3}$ corresponding to $\Ts/\Tc=1.09$ and $\Ts/\Tc=2.59$.
	(g) and (h) show results for the 2d Yukawa ABP model at states with $v_0=10$ and $v_0=50$ corresponding to $\Ts/\Tc=1.13$ and $\Ts/\Tc=2.18$.}
\end{figure}

By reference to the data in \fig{fig2}, \fig{fig2ex} compares the RDF of states predicted to be close to and not close to thermal equilibrium. Each subfigure reports $\Ts/\Tc$; results for the cases where $\Ts/\Tc$ is close to unity are found in the left column. The RDFs are compared to the equilibrium RDF for $T=\Ts$, i.e., the temperature corresponding to the potential energy of the active-matter configurations. The black dashed lines give the equilibrium RDF, the red curves are the active-matter RDFs. \Fig{fig2ex}(a)-(d) show data for $\textrm{RDF}_\textrm{AA}$ and $\textrm{RDF}_\textrm{BB}$ of the Kob-Andersen AOUP model studied in Paper I; $\textrm{RDF}_\textrm{AB}$ is similar to the AA (data not shown). \Fig{fig2ex}(e) and (f) give data for the 2d Yukawa AOUP model, while (g) and (h) give data for the 2d Yukawa ABP model (\sect{III}). \Fig{fig2ex} confirms that when the ratio $\Ts/\Tc$ is close to unity, the configurations of the active-matter model are close to thermal equilibrium configurations.

\begin{figure}[h]
	\includegraphics[width=6cm]{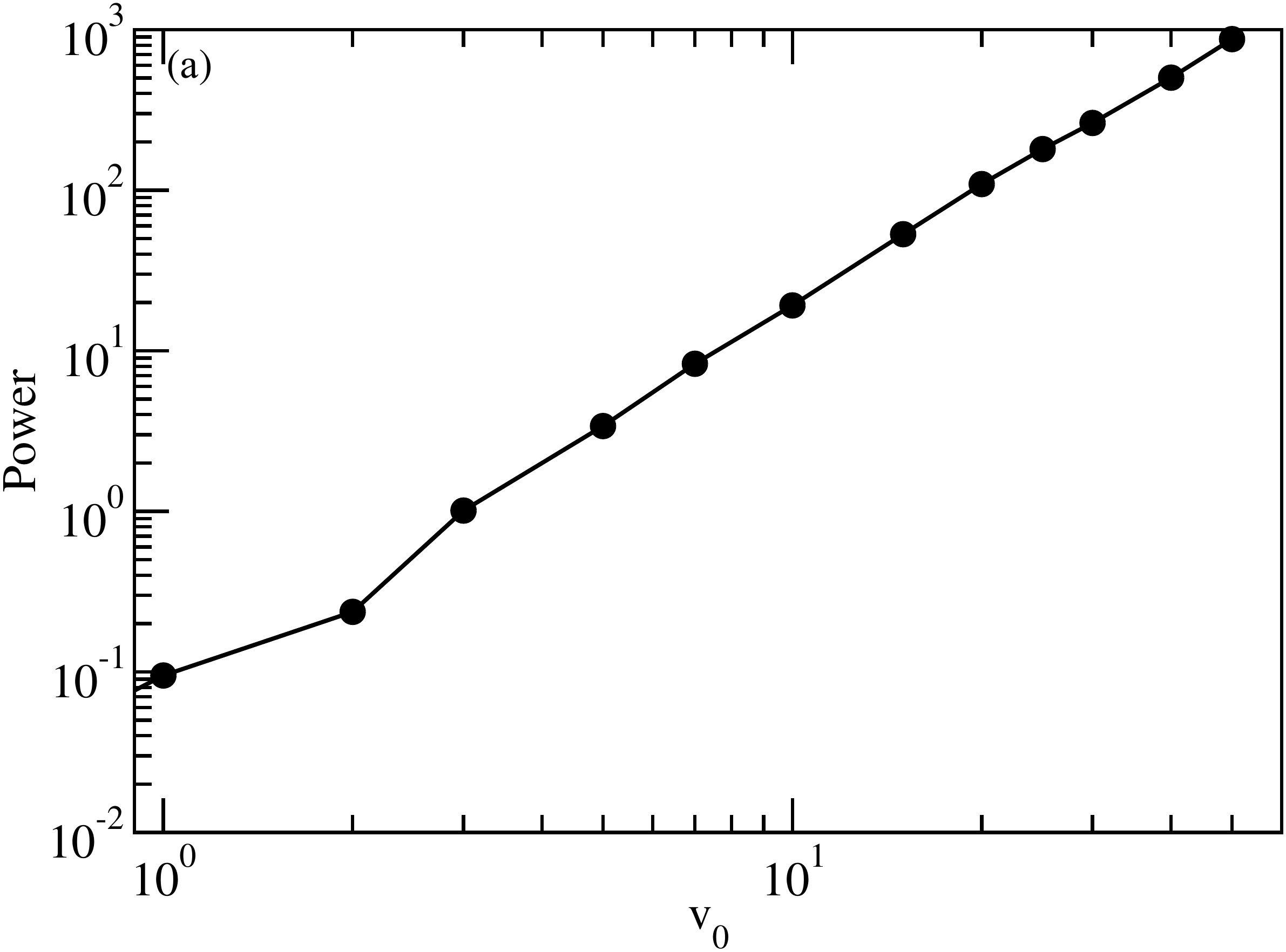}
	\includegraphics[width=6cm]{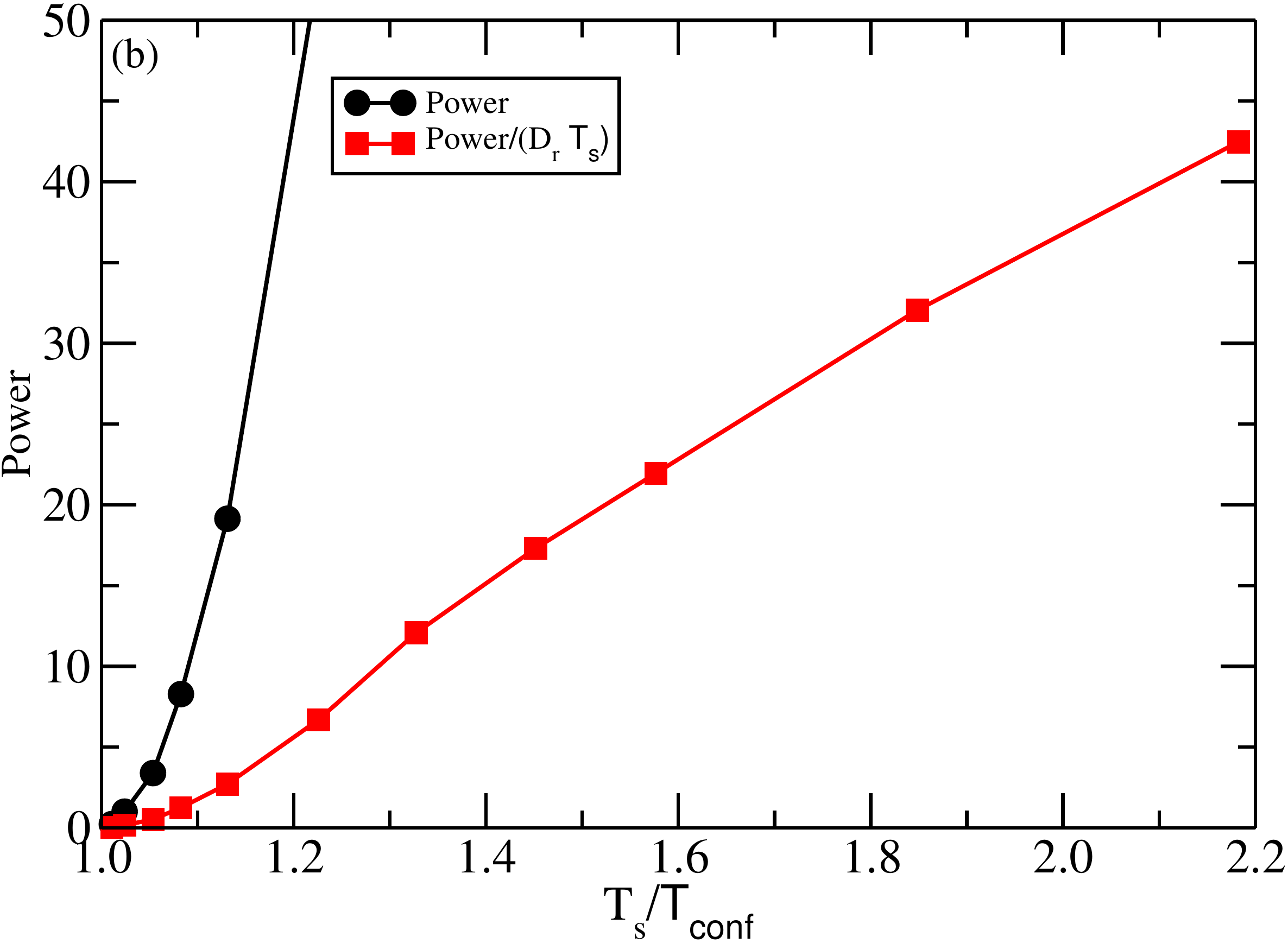}
	\includegraphics[width=6cm]{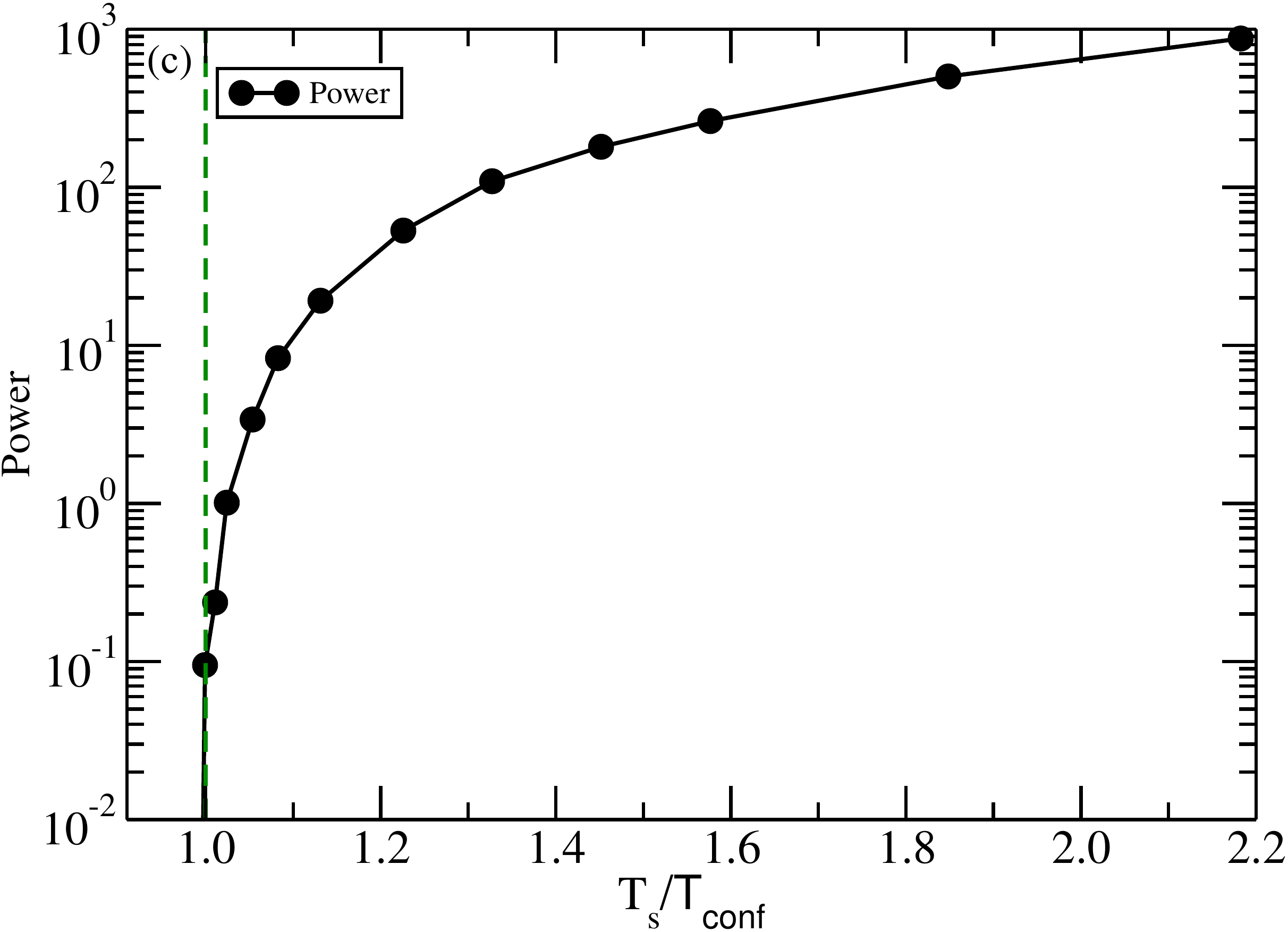}
	\caption{\label{fig3} Using the ratio of systemic to configurational temperature to quantify how far the 2d Yukawa ABP system is from thermal equilibrium (corresponding to $v_0=0$ in \eq{ABP_EOM2}); the parameters kept fixed here are $\rho=1$, $D_r=3$, and $D_t=2$. (a) shows how the dissipation (``Power'') varies with $v_0$ (MD units). From \fig{fig2}(e) we see that when $v_0\to 0$, the two temperatures become identical (equal to $2$ because $D_t=2$ corresponds to that thermal equilibrium temperature); at the same time the dissipation goes to zero. (b) and (c) show the power as a function of $\Ts/\Tc$. The quantity $\Ts/\Tc$ goes to unity as thermal equilibrium is approached, which presents an advantage compared to using the dissipated power for quantifying deviations from thermal equilibrium.}
\end{figure}

Next we compare to a previously proposed measure of deviations from thermal equilibrium, focusing on the 2d Yukawa ABP model. \Fig{fig3}(a) shows the dissipated ``active'' power, i.e., the average of the scalar product of the particle velocity with the ${v}_0\, \bo_i(t)$ term of \eq{ABP_EOM2}, plotted as a function of $v_0$, keeping the three other model parameters constant. From data like these one cannot easily determine when the system is expected to be close to thermal equilibrium. \Fig{fig3}(b) shows the dissipated power plotted against $\Ts/\Tc$, demonstrating a one-to-one correspondence between the two measures of deviations from thermal equilibrium. \Fig{fig3}(b) also includes data for the reduced-unit power (red points), which shows an interesting almost linear proportionality to $\Ts/\Tc-1$ for which we have no good explanation. Finally, \Fig{fig3}(c) plots the same data in a log-linear scale, which further illustrates that measuring deviations from thermal equilibrium in terms of a quantity that is zero in equilibrium is not useful for distinguishing between weak and stronger deviations from equilibrium.

\section{The MIPS boundary of the 2d ABP Yukawa model}\label{MIPS}

\begin{figure}[h]
	\includegraphics[width=6cm]{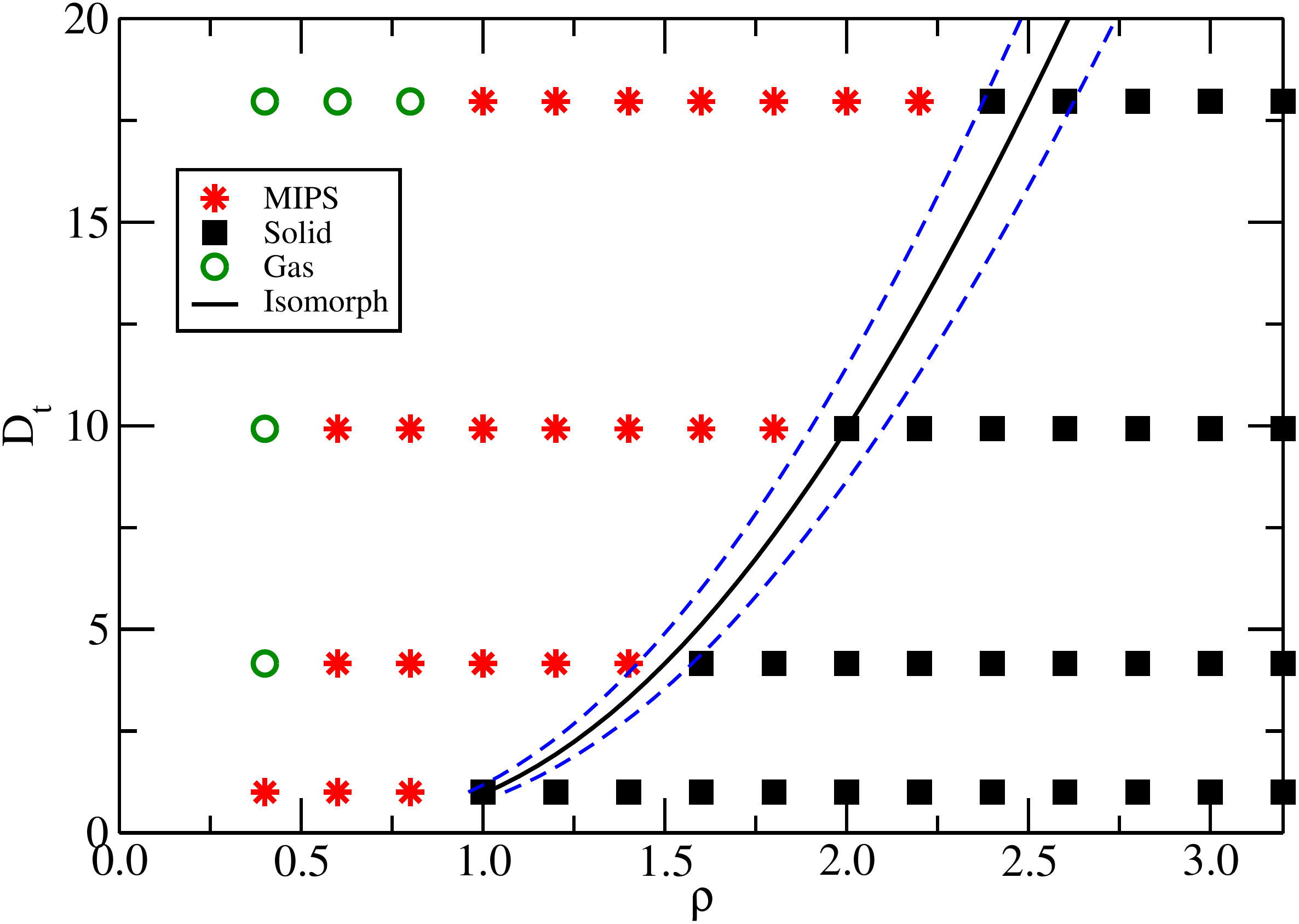}
	\caption{\label{fig4} $(\rho,D_t)$ phase diagrams showing MIPS state points as red stars and homogeneous state points as black squares (green circles are gas-like states of minor relevance here). The MIPS phase consists of coexisting phases that differ in density, the denser phase is a ``solid'' phase of hexagonal crystal structure. The reference state point $(\rho,D_r,D_t,v_0)=(1.01, 3, 1, 367)$ is located in the homogeneous (solid) phase close to the phase boundary. From this an  active-matter isomorph was traced out using \eq{ABP_param2} (black line). The figure gives data in the $(\rho,D_t)$ phase diagram with $D_r$ and $v_0$ given by \eq{ABP_param2} at density $\rho$. The blue dashed lines mark $\pm 5$\% variations in density. We see that the phase-transition line is an approximate active-matter isomorph, which is consistent with the degree of deviation from thermal equilibrium being constant along this line.}
\end{figure}

For certain parameters of the 2d ABP Yukawa model, motility-induced phase separation (MIPS) is observed. This is the striking active-matter phenomenon that even a purely repulsive system may phase separate into high- and low-density phases \cite{vic95,das14,cat15,ram17,gey19,das20,mer20}. It is reasonable to assume that, when the phase transition is approached from the homogeneous phase, the deviations from thermal equilibrium are the same for all parameter values. Thus if $\Ts/\Tc$ indeed provides a measure of the deviation from equilibrium, this quantity should be roughly constant approaching the MIPS phase transition. Since the 2d Yukawa ABP model obeys hidden scale invariance, this means that the phase transition should approximately follow an isomorph (because the physics is approximately invariant along an active-matter isomorph, such a curve cannot cross the MIPS boundary, compare Refs. \onlinecite{IV} and \onlinecite{cos16,ped16}). Thus if one has identified a state point in the homogeneous solid phase close to the MIPS boundary and uses this as reference state point for generating an active-matter isomorph, all state points identified by \eq{ABP_param2} should be close to the MIPS boundary. A similar line of reasoning was validated for the melting line of the ordinary Lennard-Jones system \cite{cos16,ped16}.

\begin{figure}[h!]
	\includegraphics[width=6cm]{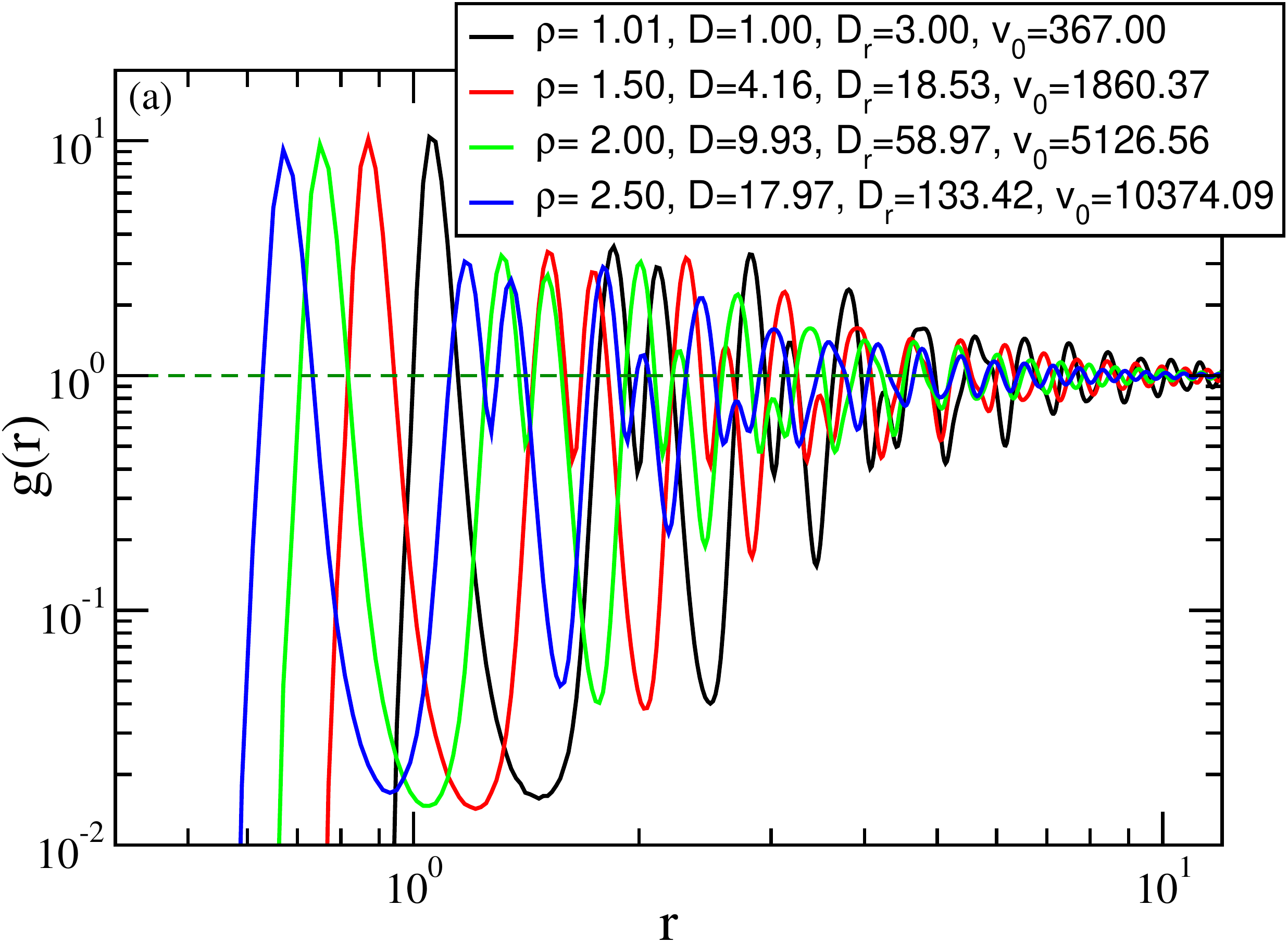}
	\includegraphics[width=6cm]{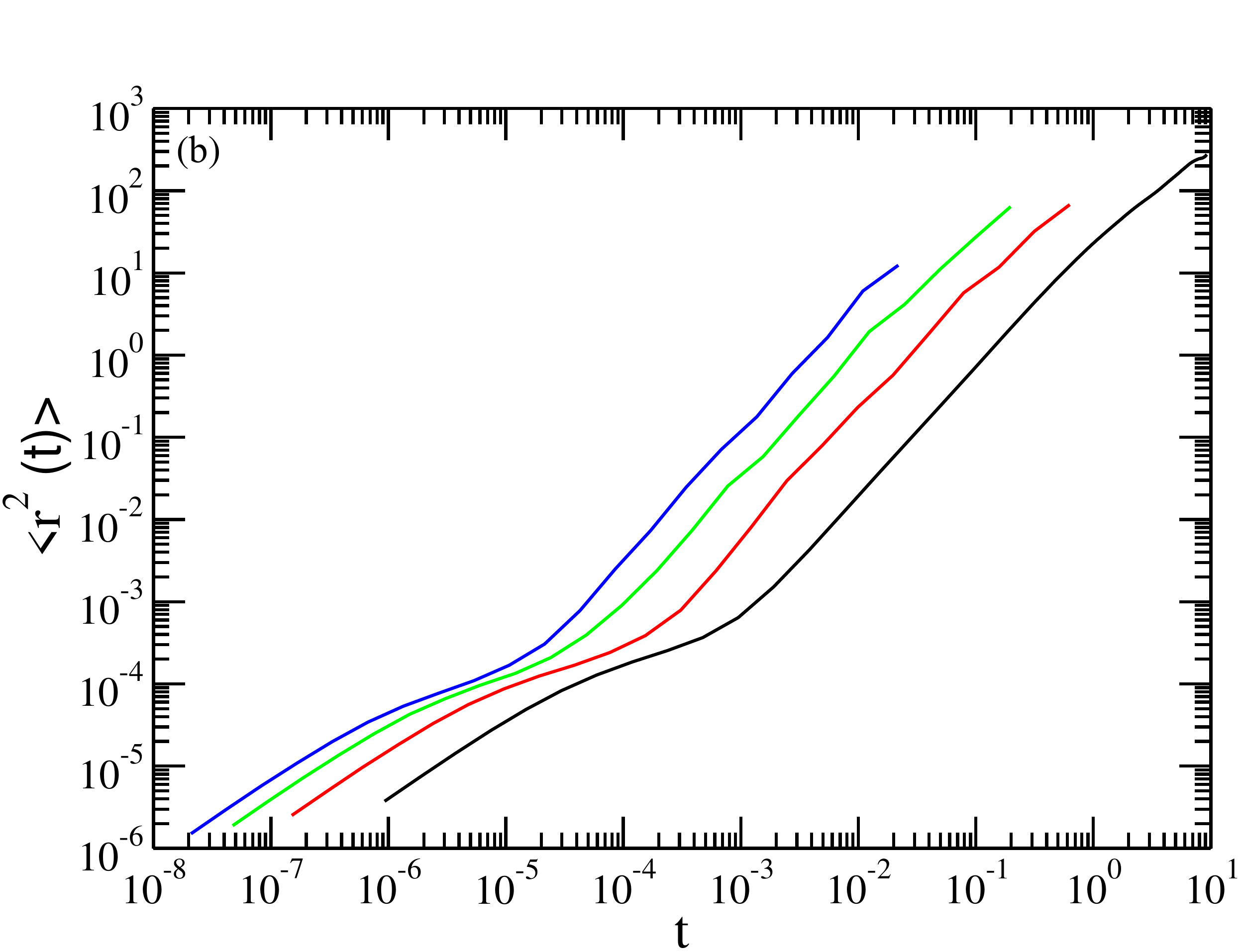}	
	\includegraphics[width=6cm]{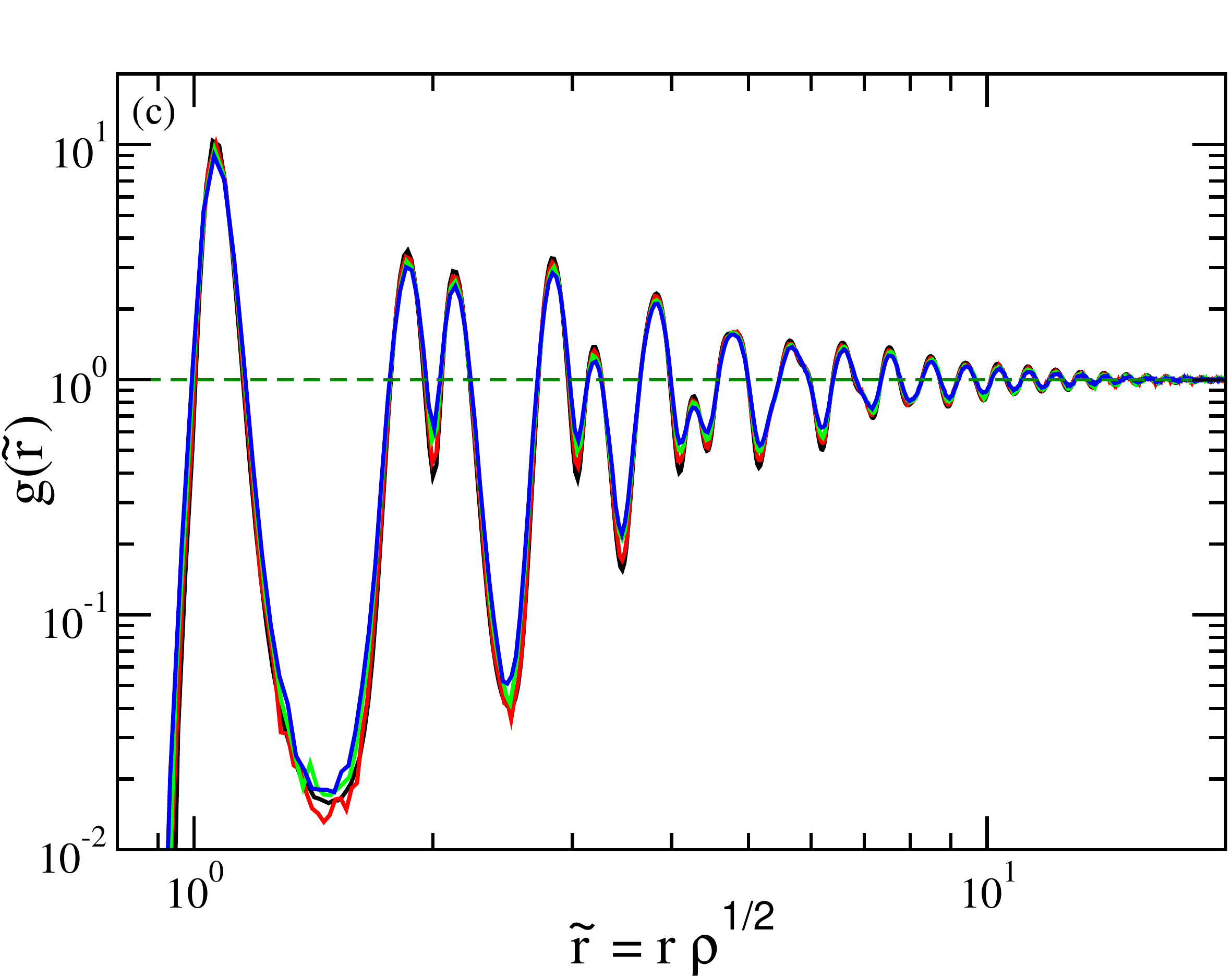}
	\includegraphics[width=6cm]{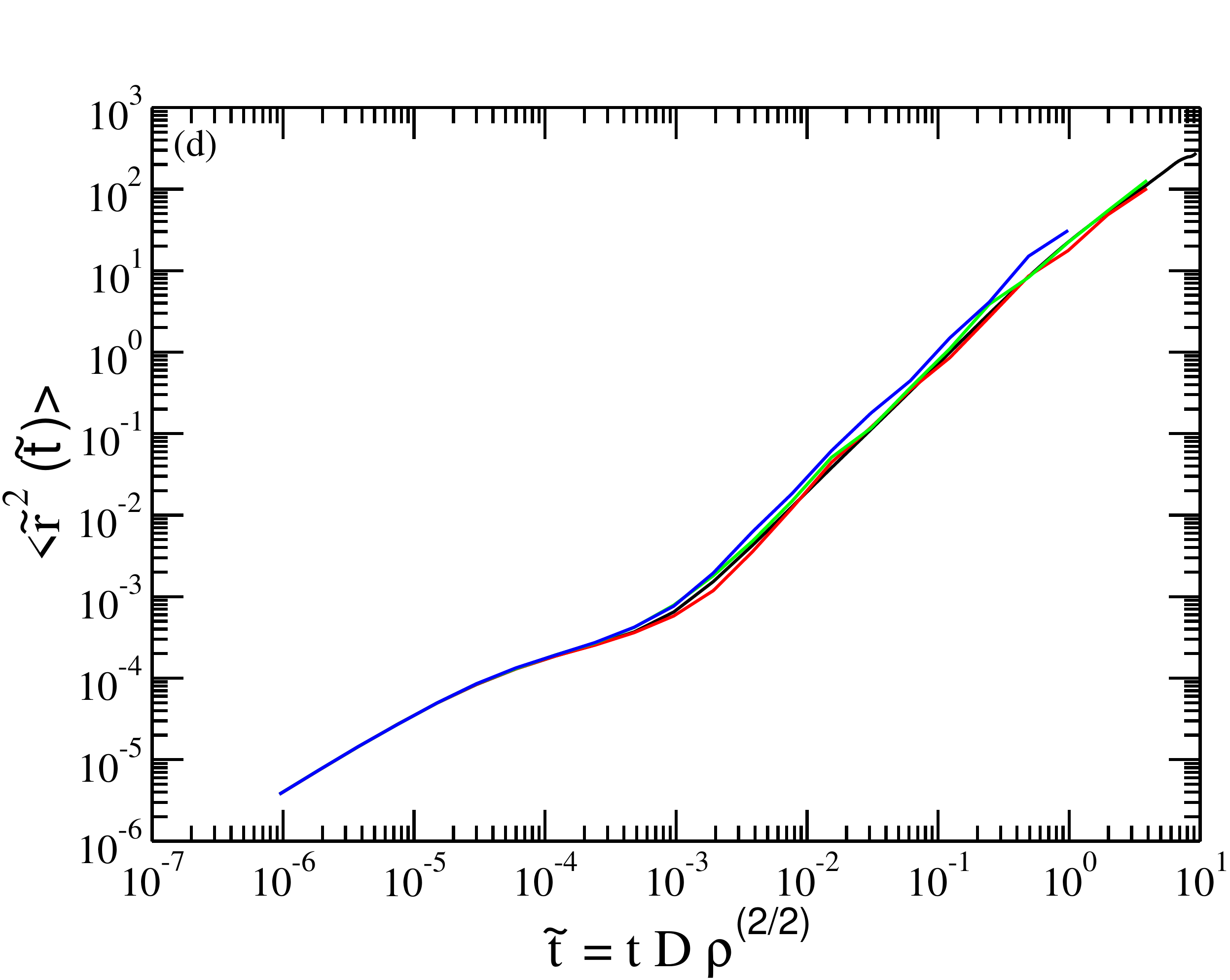}
	\caption{\label{fig5} Structure and dynamics probed along the active-matter isomorph approximately delimiting the MIPS phase boundary of the 2d ABP Yukawa system, slightly into the homogeneous phase (\fig{fig4}). (a) and (b) show log-log plots of the RDF and MSD, respectively, (c) and (d) show the same data in reduced units. }
\end{figure}

We studied the 2d Yukawa model with parameters $Q=1000$ and $\lambda=0.12$ with a cutoff at $4.2\sigma$ and $(D_r,D_t,v_0)=(3, 1, 367)$, by systematically decreasing the density from a high value well within the homogeneous solid phase. Initially, a system of 40000 particles was simulated for 40 million time steps, and the occurrence of MIPS was detected by visual inspection. The lowest density before observing MIPS was $\rho=1.01$. We then used  $(\rho,D_r,D_t,v_0)=(1.01, 3, 1, 367)$ as reference state point for generating an active-matter isomorph according to \eq{ABP_param2}. This is the black full line in \fig{fig4}, which shows the results of investigating the existence of MIPS in a $(\rho,D_t)$ phase diagram (along the isomorph the remaining parameters $D_r(\rho)$ and $v_0(\rho)$ are given by \eq{ABP_param2}). The black squares denote state points of the homogeneous solid phase, the red stars denote state points where MIPS appears, and the green circles denote gas-phase state points. The blue dashed lines mark the active-matter isomorph $\pm$5\% in density. We see that the phase transition line is predicted reasonably well though not accurately; this is consistent with the approximate nature of the argument. Nevertheless, the simulations demonstrate that \eq{ABP_param2} can be used for roughly identifying the MIPS phase boundary. This confirms the physical expectation that the deviation from thermal equilibrium is virtually constant along the phase-transition line because it is an approximate active-matter isomorph characterized by constant $\Ts/\Tc$.

In order to confirm that the black line of \fig{fig4} is a line of approximately invariant physics, i.e., an active-matter isomorph, we show in \fig{fig5} how structure and dynamics vary along it. The upper figures show the RDF and MSD in standard units, the lower figures show the same data in reduced units.

\section{Summary of Papers I \& II and Outlook}\label{out}

The configurational-temperature concept has traditionally been used in connection with liquid models based on Newton's laws of motion with forces derived from a potential-energy function $U(\bR)$ \cite{pow05}. Indeed, the derivation of $\Tc$ refers to the canonical ensemble, and for this reason it is not obvious that $\Tc$ has relevance also for non-Hamiltonian and non-time-reversible systems like those of active matter. We have suggested that the configurational temperature may be useful also in that context and have presented two applications of $\Tc$. Paper I demonstrated how $\Tc$ may be used for tracing out lines of approximately invariant structure and dynamics in the phase diagram of models described by AOUP dynamics if the potential-energy function obeys hidden scale invariance; such lines are referred to as active-matter isomorphs. Specifically, Paper I gave the equations for how to change the model parameters with density in order to have invariant physics, and Paper II derived a similar procedure for ABP models. In both cases, by effectively reducing the number of model parameters by one, this approach provides a tool for simplifying the exploration of phase diagrams of active-matter models with hidden scale invariance of the potential-energy function.

For the AOUP and the ABP models the ratio of systemic to configurational temperature is predicted to be constant along an active-matter isomorph. Since both the active-matter physics and the corresponding passive-matter physics are invariant along their common systemic isomorph (defined as the thermal equilibrium isomorph mapped into the density systemic-temperature phase diagram \cite{dyr20}), this is consistent with the present paper's proposal that $\Ts/\Tc$ quantifies how far a given active-matter system is from thermal equilibrium. 

The ratio $\Ts/\Tc$ is defined for any active-matter system based on a potential-energy function, whether or not hidden scale invariance applies. We suggest that an active-matter system may be regarded as ``close to thermal equilibrium'' whenever $\Ts/\Tc$ is close to unity and ``far from thermal equilibrium'' whenever this is not the case. We illustrated the use of $\Ts/\Tc$ for quantifying deviations from thermal equilibrium by showing that when this quantity is close to unity, the RDF of the active-matter system is close to that of the corresponding thermal-equilibrium system with $T=\Ts$. Moreover, $\Ts/\Tc$ is roughly constant along the motility-induced phase separation (MIPS) boundary along which the deviation from equilibrium are expected not to vary, compare \fig{fig5}. 

The advantages of using the quantity $\Ts/\Tc$ for quantifying how far an active-matter system is from thermal equilibrium are threefold:

\begin{itemize}
\item A measure that converges to unity when the system in question approaches thermal equilibrium allows for answering the question: how to quantify the deviation from thermal equilibrium? This is not the case for a measure that converges to zero when equilibrium is approached.
\item $\Ts/\Tc$ is easy to evaluate because it can be determined from a single configuration $\bR$ of a steady-state simulation of the active-matter system in conjunction with equilibrium simulations of the corresponding Hamiltonian system.
\item $\Ts/\Tc$ is general measure because this quantity is defined for any system characterized by a potential-energy function, whether or not in the context of an active-matter model. For instance, in the case of a non-linear steady-state shear flow of an ordinary Hamiltonian system, it is also possible to quantify the deviation from thermal equilibrium by means of $\Ts/\Tc$. 
\end{itemize}
An interesting question that remains to be explored is the following: What is the difference between the cases $\Ts/\Tc>1$ and $\Ts/\Tc<1$?

\begin{acknowledgments}
	This work was supported by the VILLUM Foundation's \textit{Matter} grant (16515).
\end{acknowledgments}

\end{document}